\documentclass[letterpaper, reprint, superscriptaddress, amsmath, amssymb, aps, pre, nofootinbib]{revtex4-1}

\usepackage{graphicx}
\usepackage{dcolumn}
\usepackage{bm}
\usepackage{hyperref}
\usepackage{subfigure}  
\usepackage{xcolor}  
\usepackage[normalem]{ulem} 

\newcommand{\Mmax}{M_{\textrm{max}}} 
\newcommand{\Mtot}{M_{\textrm{tot}}} 
\newcommand{\Mcrit}{M_{\textrm{crit}}} 
\newcommand{\re}[1]{\textrm{Re}\left(#1\right)} 
\newcommand{\im}[1]{\textrm{Im}\left(#1\right)} 
\newcommand{\avg}[1]{\left\langle#1\right\rangle} 
\newcommand{\paren}[1]{\left(#1\right)} 

\graphicspath{{Figures/}}

\begin{document}

\preprint{APS/123-QED}

\title{Symmetry-broken states on networks of coupled oscillators}%

\author{Xin Jiang}
\email[E-mail me at: ]{jiangxin@buaa.edu.cn}
\affiliation{Department of Engineering Sciences and Applied Mathematics, Northwestern University, Evanston, Illinois 60208, USA}
\affiliation{LMIB and School of Mathematics and Systems Science, Beihang University, Beijing 100191, China}
\author{Daniel M.~Abrams}
\affiliation{Department of Engineering Sciences and Applied Mathematics, Northwestern University, Evanston, Illinois 60208, USA}
\affiliation{Northwestern Institute for Complex Systems, Northwestern University, Evanston, Illinois 60208, USA}
\affiliation{Department of Physics and Astronomy, Northwestern University, Evanston, Illinois 60208, USA}


\begin{abstract}
When identical oscillators are coupled together in a network, dynamical steady states are often assumed to reflect network symmetries.  Here we show that alternative persistent states may also exist that break the symmetries of the underlying coupling network.  We further show that these symmetry-broken coexistent states are analogous to those dubbed ``chimera states,'' which can occur when identical oscillators are coupled to one another in identical ways.
\end{abstract}

\maketitle

When identical oscillators are coupled to one another in identical ways, it is unsurprising that synchronization may occur.  However, another apparently stable symmetry-broken state of partial synchrony may also occur; this has come to be known as a ``chimera state'' \cite{Kuramoto2002,Abrams2004,Panaggio2015_2}.  In this paper we show that chimera states are robust to significant changes in the network structure, and that the resulting irregular networks can retain apparently stable states that bear the same symmetry-breaking hallmark as the chimera state from the original network.

\begin{figure}[t!]
  \centering
  \begin{minipage}[c]{0.49\columnwidth}%
    \hspace{5mm}
    \includegraphics[width=0.75\columnwidth]{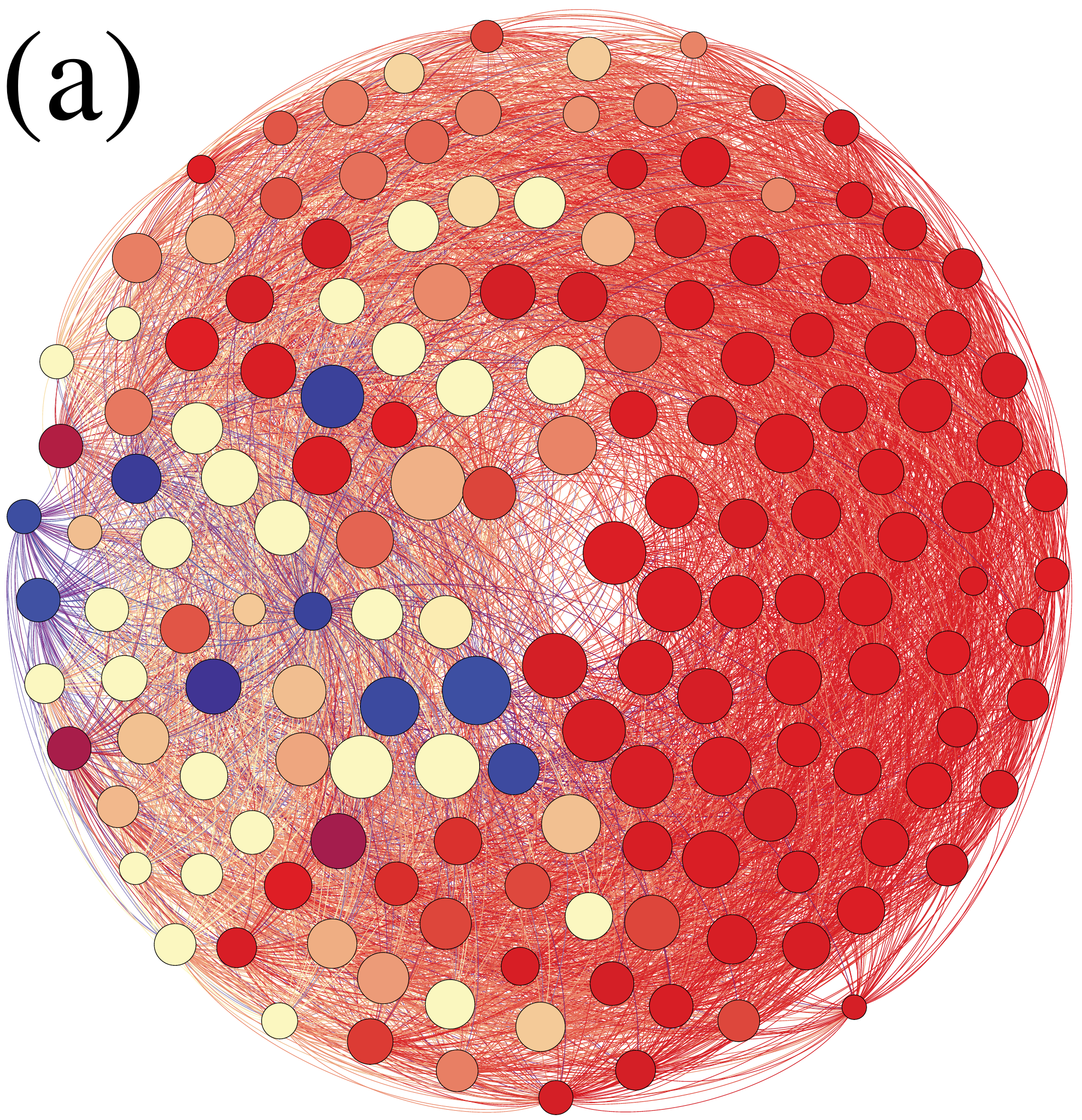}
  \end{minipage}
  \begin{minipage}[c]{0.49\columnwidth}%
    \hspace{1mm}
    \includegraphics[width=0.75\columnwidth]{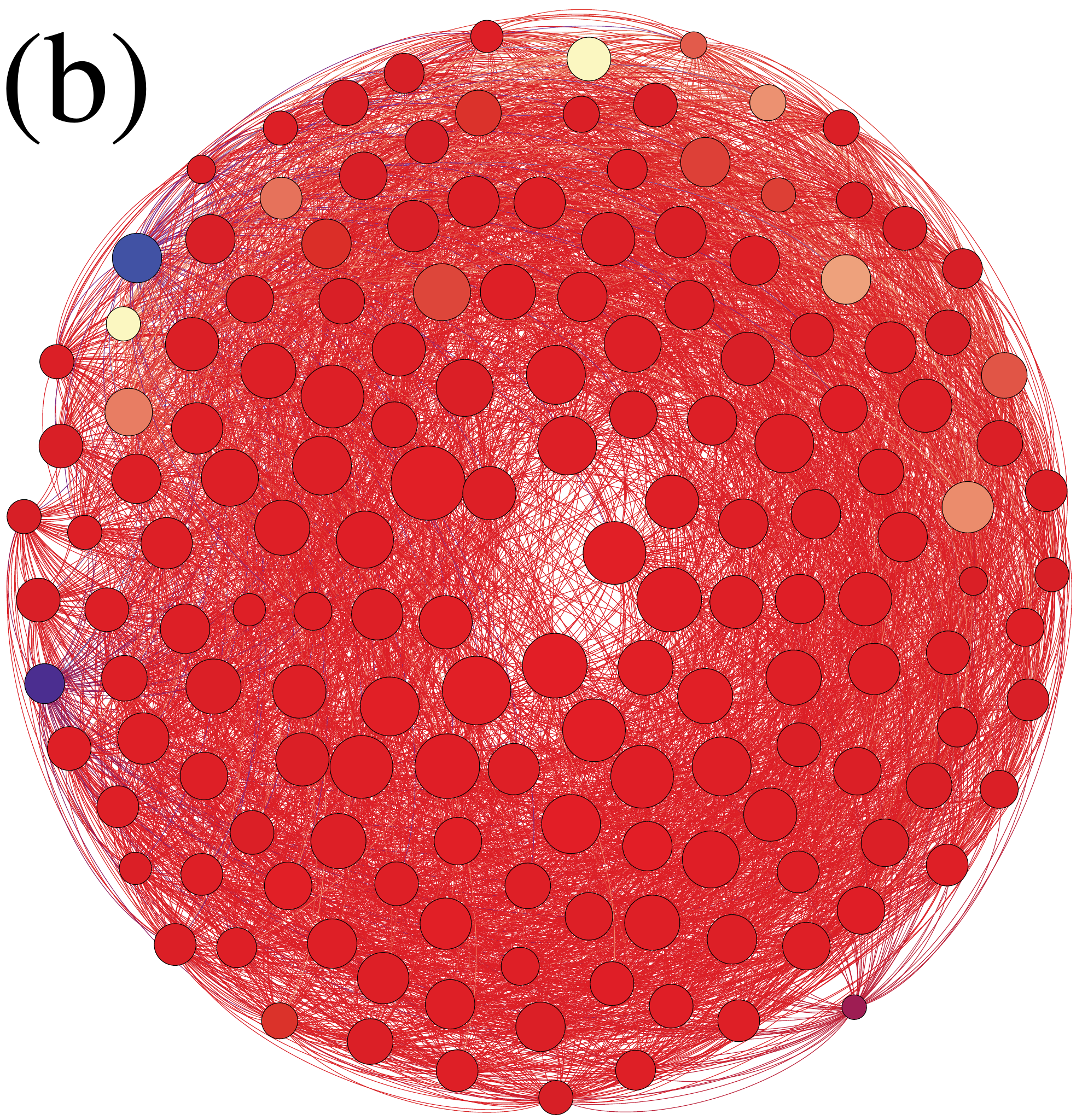}
  \end{minipage}
  \begin{minipage}[c]{\columnwidth}%
    \vspace{3mm}
    \includegraphics[width=0.95\columnwidth,clip]{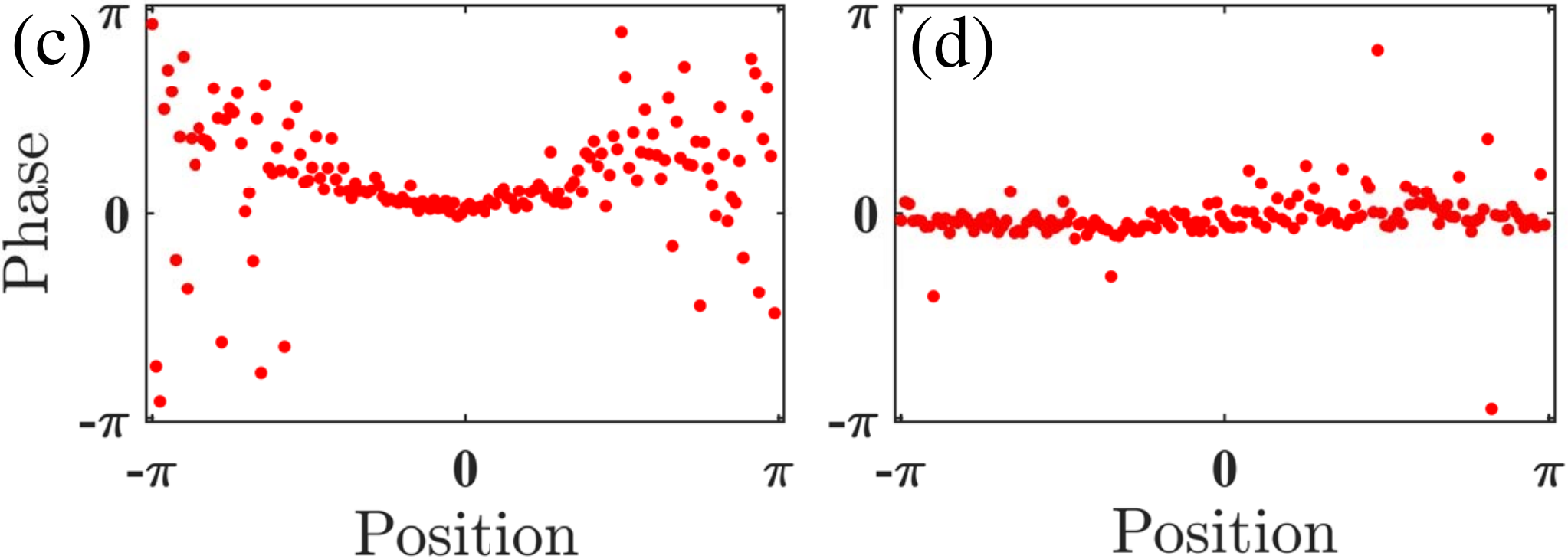}
  \end{minipage}
  \caption{(Color online)\textbf{Two different stable states on a network}.  Upper left: Phase pattern for a partially-synchronized state observed on a network. Upper right: Phase pattern for a fully-synchronized state observed on the same network. In both panels node size indicates degree and color indicates phase $\phi$ (white-red-blue as phase varies from $-\pi$ to $\pi$). Lower panels: phase patterns from upper panels showing phase versus node spatial position $x_i$ (node spatial position is given by node angle in upper panels; radial position has no significance). For this network $N=160$ oscillators and $\alpha=1.34$, with dynamics according to Eq.~\eqref{eq:ode}.}
  \label{fig:twostates}
\end{figure}

Figure \ref{fig:twostates} shows two different equilibria, both apparently stable, on a single irregular network consisting of 160 oscillators.  The right panels show a state that is roughly spatially invariant, and the left panels a symmetry-broken state analogous to the chimera state.  An important question for real-world networks is whether the existence of non-synchronous steady states can be predicted from the network structure \cite{Pecora2014, Lin2015, Fu2013, Zhou2006, Belykh2008, Nicosia2013, Arenas2008, Laing2009_2, Timme2004, Rodrigues2016}.  This may have implications for systems ranging from generators on an electrical power grid \cite{Motter2013, Menck2014, Rohden2014, Nishikawa2015} to cardiac pacemaker cells in the heart \cite{Dehaan1972, Glass2001, Taylor2015, Skardal2014} to the neurons in the suprachiasmiatic nucleus of the brain responsible for circadian rhythms \cite{Yamaguchi2003, Gu2012, Stoleru2004, Laing2009_1, Hizanidis2014}.

\paragraph*{Generalizing chimera states.}
We wish to generalize the symmetry-breaking phenomenon responsible for the existence of chimera states to randomly connected oscillator networks. We accomplish this algorithmically by making incremental perturbations to a symmetric network on which chimera states have been observed and explained in the continuum limit: the one-dimensional ring of oscillators \cite{Abrams2004,Abrams2006,Omelchenko2013}.

By randomly rewiring links while keeping the total number constant, we can gradually tune the network from spatially structured (in an ensemble average sense) to an Erd\"{o}s-R\'{e}nyi (ER) random network.  We perform this rewiring in an ``adiabatic'' sense, allowing the dynamical system to relax to equilibrium after each change to the network structure.

We find that bistable coexistence of a synchronous state and a chimera state persists for a significant degree of network rewiring.  Furthermore, this persistence can be explained analytically by examining both the dynamical system and the expected network structure in the thermodynamic $N \to \infty$ limit.

\paragraph*{Numerical exploration.}
We consider $N$ coupled oscillators assigned uniformly spaced positions $x_i \in (-\pi, \pi]$, $i =1 \ldots N$ along a one-dimensional ring with periodic boundary conditions. We numerically implement the following governing equation for the system:
\begin{equation}
  \label{eq:ode}
  \frac{\partial \phi_i}{\partial t}= \omega - \frac{1}{N}\sum_{j=1}^{N} c_{ij} \sin [\phi_i(t)-\phi_j(t)+\alpha],
\end{equation}
where $\omega$ is the (identical) natural frequency of each oscillator, $\phi_i(t)$ is the phase of the oscillator at position $x_i$ at time $t$, $\alpha$ is a ``phase lag'' parameter, and $c_{ij}$ is the $ij$th element of the $N \times N$ coupling matrix $\mathbf{C}$. We first consider the case of binary coupling where $c_{ij}=1$ if there is an undirected link between oscillators $i$ and $j$, and $c_{ij}=0$ otherwise.

In the thermodynamic limit $N \to \infty$, phase becomes a continuous function of both space and time $\phi(x,t)$ and the adjacency matrix $\mathbf{C}$ must be replaced by a continuous coupling kernel $G(x,x')$.  Such a system was studied in ref.~\cite{Abrams2004} with
\begin{equation}
  \label{eq:chimerakernel}
  G(x,x') = \frac{1}{2\pi}[1+A \cos(|x-x'|)],
\end{equation}
where parameter $A$ tunes the locality of the coupling. It was found that chimera states could stably exist for certain ranges of the parameters $\alpha$ and $A$.

\begin{figure}[t!]
  \centerline{
  \includegraphics[width=\columnwidth]{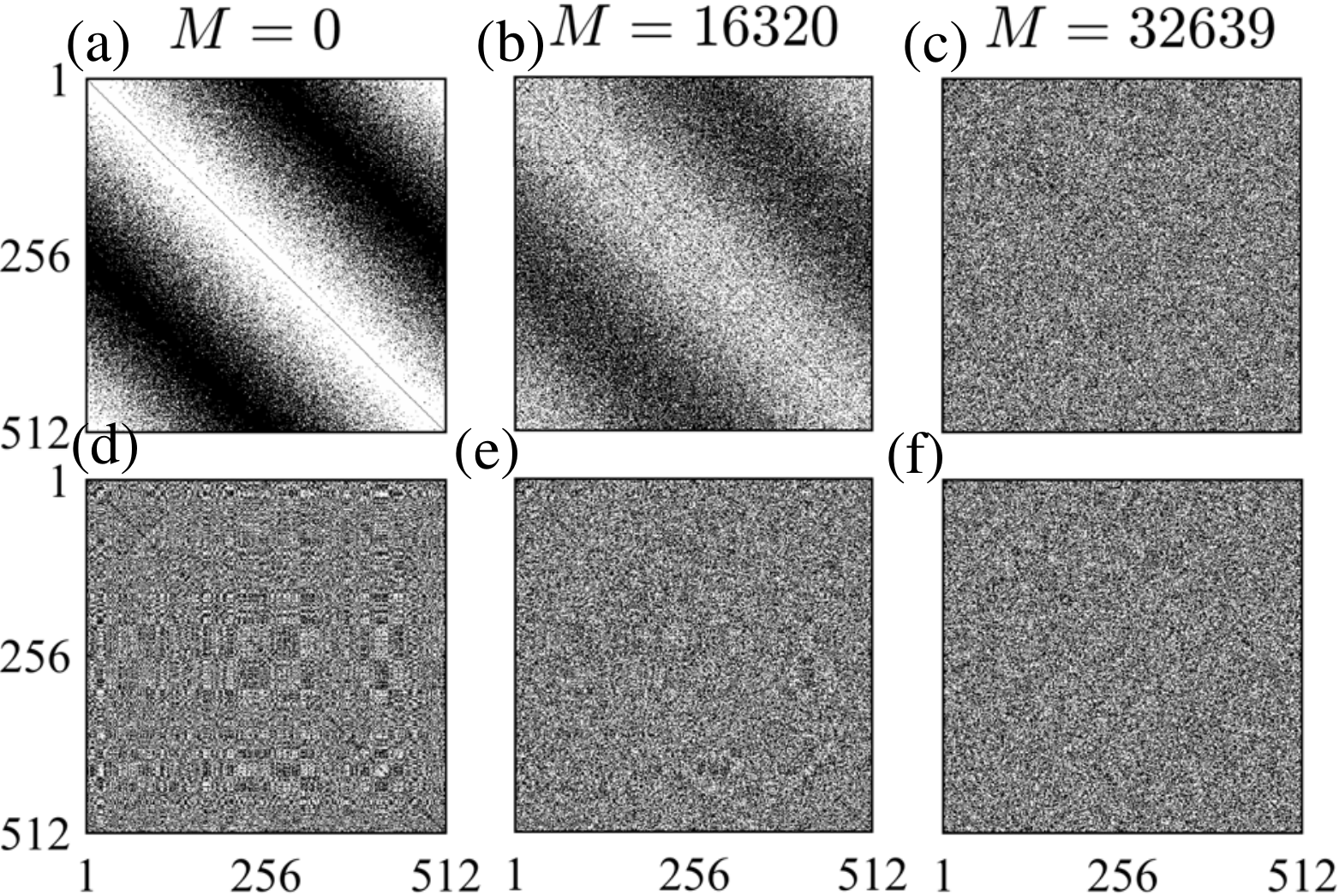}}
  \caption{\textbf{Adjacency matrices}. The adjacency matrix for the initial network (left top panel), a network with $M=16320$ links rewired (middle top panel), and the Erd{\H{o}}s-R{\'e}nyi random target network with $\Mmax = 32639$ links rewired (right top panel, $\Mtot=65383$).  Bottom panels display the same three networks but with different node enumeration, emphasizing the difficulty in recognizing spatial network structure without a-priori knowledge of its existence. White indicates presence of a link, black indicates absence of a link. Here $N=512$, $\lambda=\pi$, $A=1$, and self-coupling is excluded.}
  \label{fig:rewiring_matrix}
\end{figure}

For the purpose of numerical simulation, finite sized networks are required.  We generate a random adjacency matrix $\mathbf{C}$ where, motivated by Eq.~\eqref{eq:chimerakernel}, the probability of connection between nodes $i$ and $j$ is given by
\begin{equation}
  \label{eq:linkprob}
  p_{ij}=\frac{\lambda}{2\pi}[1+A \cos(|x_i-x_j|)].
\end{equation}
Here $\lambda$ is a parameter that can be used to tune the link density (average degree) of the network, and $0 \le A \le 1$, and $0 \le \lambda \le 2 \pi / (1+A)$ is assumed in order to keep probabilities in allowed ranges.  See the upper left panel of Fig.~\ref{fig:rewiring_matrix} for an example.  We denote the total number of links in a given network realization $\Mtot$, with expected valued approximately $\lambda N^2 / (4 \pi)$.

After generating our initial structured random network according to Eq.~\eqref{eq:linkprob}, we use a variant of the Erd{\H{o}}s R\'{e}nyi (ER) random graph model \cite{Erdos1960,Newman2003} to generate a ``target'' network with an identical number of nodes and links (see upper right panel of Fig.~\ref{fig:rewiring_matrix}). We refer to the total number of links that differ between the initial network and the target network---the maximum number of links that can be rewired---as $\Mmax$. We then rewire our initial network in a step-by-step fashion: we randomly choose a link that exists in the initial network but not in the target network, delete it and replace it with one that exists in the target network but not in the initial network, then let the dynamical system equilibrate.  Repeating this process gradually transforms the initial network into the target ER network while minimally disturbing the dynamical steady-state.

To characterize the dynamics of the system as we proceed with this rewiring experiment, we introduce the global order parameter
\begin{equation}
  R = \left| \frac{1}{N}\sum_{j=1}^{N} e^{i\phi_{j}} \right|
  \label{eq:orderparam}
\end{equation}
and track how it changes as the system is rewired in the pseudo-static fashion described above.

For sufficiently large $N$, the initial network supports both chimera and synchronous states, both of which appear to be numerically stable\footnote{Omel'chenko \cite{Omelchenko2013} demonstrated that chimera states in systems of this type are extremely long-lived transients, but for $N \gtrsim 2^7$ the expected lifetimes are much longer than the duration of our numerical simulation.}.  We select initial conditions (ICs) from both basins of attraction for the same initial network and (in separate experiments) track how those states evolve as the rewiring proceeds (see Appendix \ref{app:num_notes} for details on the numerical approach).

\begin{figure}[t!]
  \centerline{
  \includegraphics[width=\columnwidth]{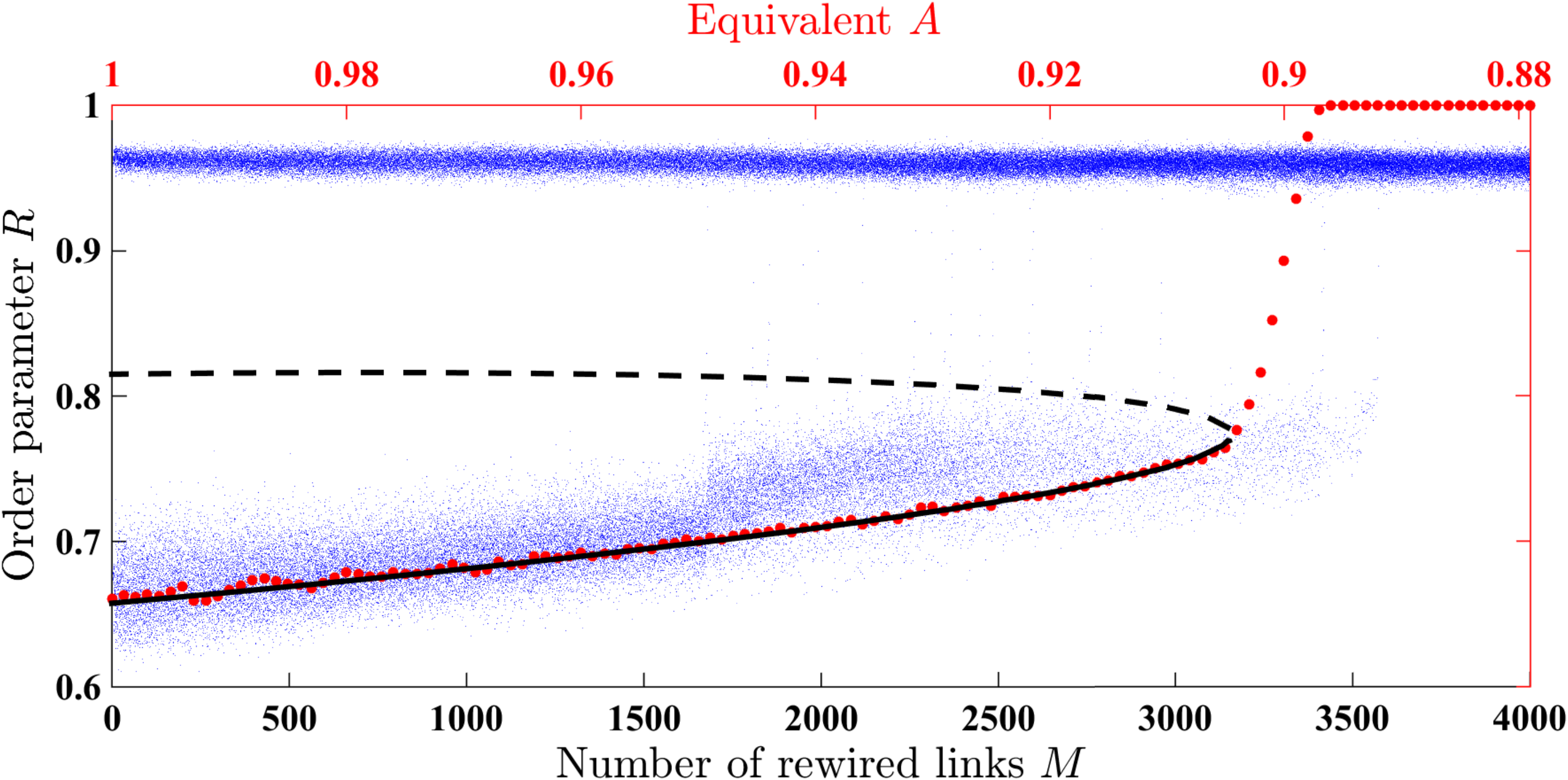}
  }
  \caption{(Color online)\textbf{Results of rewiring experiment}. Blue points: Numerical results from 20 rewiring experiments on binary networks. Red points: Averaged results from an ensemble of 50 numerical experiments on finite weighted networks. Black lines: Theoretical chimera states from model in \cite{Abrams2004} for $N \to \infty$ limit. Solid indicates stable branch, dashed indicates unstable branch. In all experiments $N=512$, $A=1$, $\lambda=\pi$, and $\alpha=1.34$.}
  \label{fig:rewiring_results}
\end{figure}

Figure \ref{fig:rewiring_results} displays the results of an ensemble of these experiments.  For initial conditions belonging to the sync basin of attraction, rewiring does not appreciably change the state of the system: the system retains high order paramer $R \approx 1$ (deviations are due to finite-size effects).  For initial conditions belonging to the chimera basin of attraction, however, something more interesting occurs.  Initially the order parameter $R$ has an intermediate value between zero and one, indicating partial synchrony, as expected.  As the number of rewired links $M$ increases, the order parameter slowly increases until it appears to suddenly jump up to $R \approx 1$ at some critical value $\Mcrit$.

\paragraph*{Connecting to theory.}
In reference \cite{Abrams2004}, a self-consistency approach is used to find solutions to the continuum version of Eq.~\eqref{eq:ode} with coupling kernel \eqref{eq:chimerakernel}.  Searching for a complex order parameter solution of the form $R(x) e^{i \Theta(x)} = c + a \cos x$, the authors come up with two complex equations for four real unknowns $c$, $\re{a}$, $\im{a}$, and $\Delta$ (the mean frequency of the synchronized oscillators).  The spatially averaged order parameter $\frac{1}{2 \pi}\int_{-\pi}^{\pi} R(x) dx = c$ is the equivalent of the global order parameter \eqref{eq:orderparam} that we track in our experiments.

\begin{figure}[t!]
  \centerline{
  \includegraphics[width=\columnwidth]{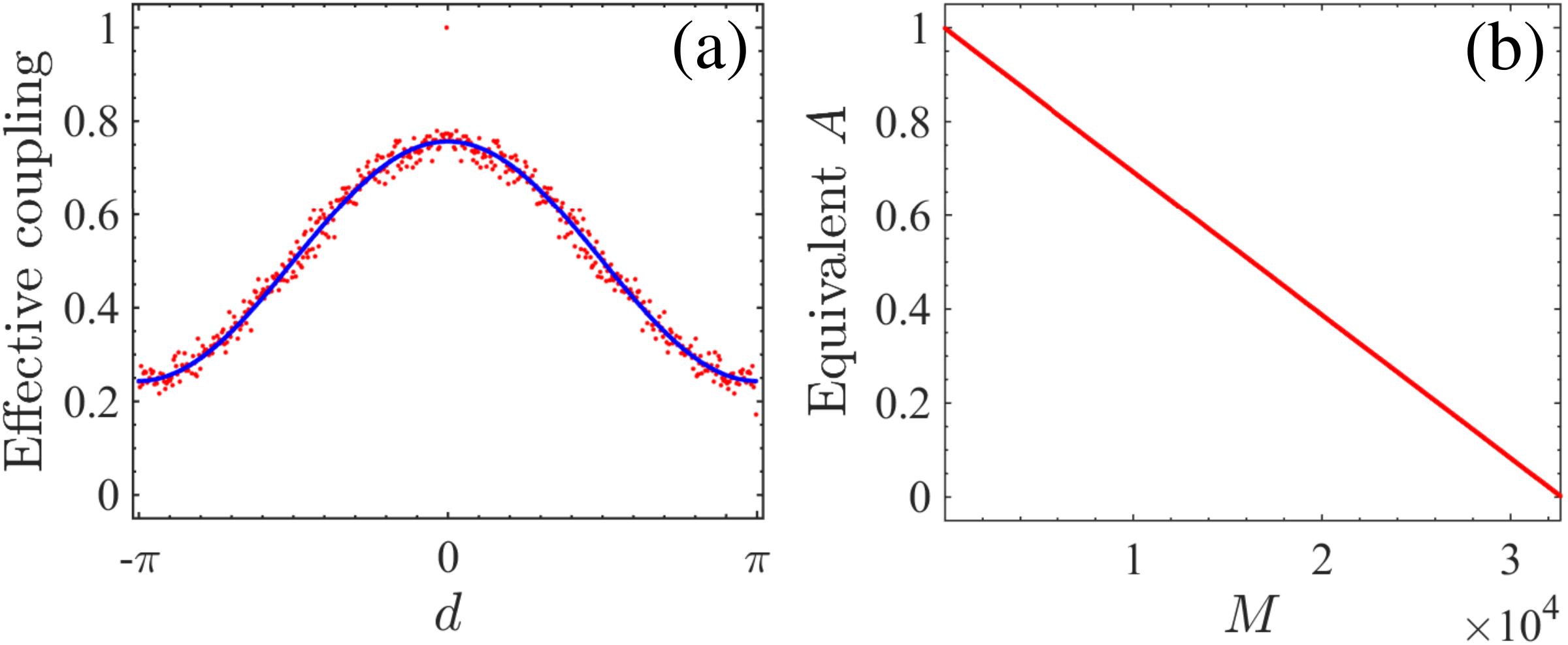}}
  \caption{(Color online) \textbf{Converting between network and continuum models}. Left panel: Red dots indicate fraction of links at distance $d$ from a given node, averaged over all nodes, for a network with $M=16000$ links rewired. Blue line shows best-fit cosine kernel coupling function $\frac{1}{2}[1 + A \cos(x)]$. Right panel: The least-squares best fit ``equivalent $A$'' versus the number of rewired links $M$ in ten binary network rewiring experiments.  Note that error bars indicating $\pm$ one standard deviation, if included, would not be visible (std.~dev.~$\sim 2 \times 10^{-3}$). Here $N=512$, $\alpha=1.34$, $\lambda=\pi$. }
  \label{fig:equivalentA}
\end{figure}

Those two complex equations,
\scriptsize
\begin{equation} \label{eq:newc}
    c = i e^{-i \alpha} \avg{ {\frac{\Delta-\paren{\Delta^2 - c^2 - 2 \re{a} c \cos x - |a|^2\cos^2 x}^{\frac{1}{2}}} {c + a^* \cos x}} }  \nonumber \\
\end{equation}
\begin{equation} \label{eq:newa}
    a = i A e^{-i \alpha} \avg{ { \frac{\Delta-\paren{\Delta^2 - c^2 - 2 \re{a} c \cos x - |a|^2\cos^2 x}^{\frac{1}{2}}} {c + a^* \cos x} \cos x} },
\end{equation}
\normalsize
can be solved numerically for $c$ as a function of $A$ to obtain theoretical predictions from the continuum limit.  This is shown as the black line in Fig.~\ref{fig:rewiring_results}, with solid indicating a stable branch of chimera-state solutions and dashed indicating an unstable branch of chimera-state solutions.  The fully synchronous state $R=1$ (not drawn) is also a theoretical solution of the continuum model.

To compare predictions of the continuum theory with results from our simulations on rewired networks, we look for a relationship between the number of links rewired $M$ and the coupling locality $A$.  As $M \to \Mmax \approx \Mtot/2$, the network becomes completely disordered, i.e., the probability of a link between any pair of nodes becomes independent of their spatial position.  This is equivalent to the limit where $A \to 0$, when coupling kernel \eqref{eq:chimerakernel} represents global coupling with no spatial structure.  Additionally, the limit $M \to 0$ represents maximal spatial order in the random network, comparable to $A \to 1$ in Eq.~\eqref{eq:chimerakernel}.

Figure \ref{fig:equivalentA}'s left panel shows a fit of the cosine kernel form \eqref{eq:chimerakernel} to data from a sample rewired network with fixed $M$.  The fraction of links connecting to a node is plotted versus spatial distance from that node, averaged over all nodes; Eq.~\eqref{eq:chimerakernel} with best-fit parameter $A$ is overlaid.  The right panel shows the resulting relationship between $A$ and $M$, plotting the least-squares best fit ``equivalent'' $A$ for each value of the rewiring number $M$ (see Appendix \ref{app:rewiring} for theoretical reasons to expect linearity between ``equivalent $A$'' and $M$ in this rewiring experiment; however, the numerical approach described here to find ``equivalent A'' should work for general coupling matrices).

When this relationship is used to connect the $A$ and $M$ scales in Fig.~\ref{fig:rewiring_results}, good agreement is seen between the continuum theory and our finite network experiments.  This strongly suggests that the chimera state phenomenon is indeed responsible for the partially synchronized steady state visible in Fig.~\ref{fig:twostates} and those indicated by the blue dots in Fig.~\ref{fig:rewiring_results}.

\begin{figure}[t!]
\centerline{
  \includegraphics[width=\columnwidth]{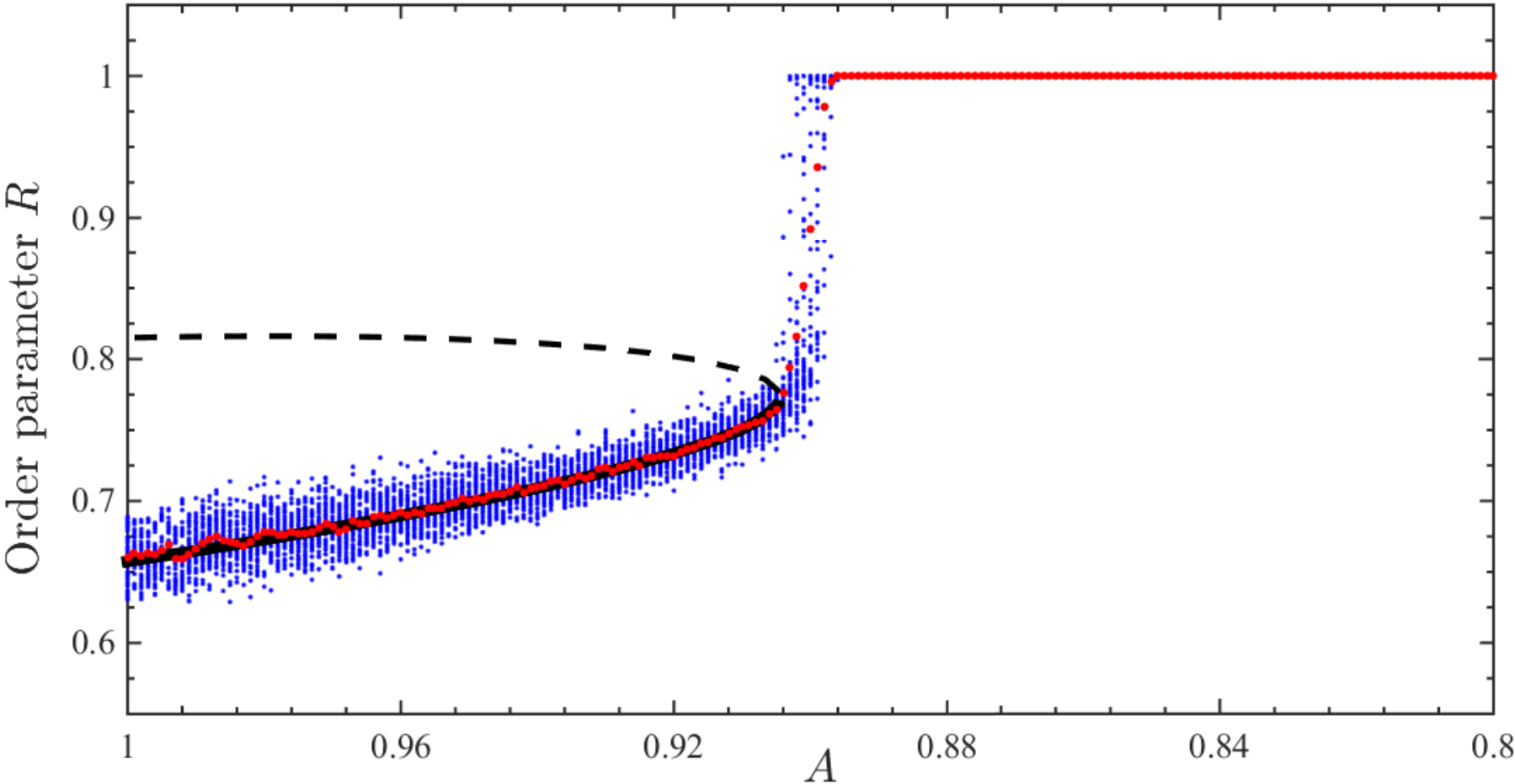}}
  \caption{(Color online) \textbf{Theory versus numerical experiment on weighted network}. Blue dots: results from an ensemble of 50 numerical experiments on the same all-to-all weighted network with different initial conditions.  For each IC, $A$ was numerically continued from $1$ down to $0.8$.  Red dots: Ensemble average order parameter value for each $A$ value.  Black curve: theory from continuum limit.  Solid line, stable; dashed line, unstable. Here $N=512$, $\alpha=1.34$, $\lambda=\pi$.}
  \label{fig:weighted_results}
\end{figure}

\paragraph*{Weighted network.}
We conducted a second set of numerical experiments, this time not restricted to binary networks but instead using all-to-all coupled networks with  weighted links. Link weights were assigned according to Eq.~\eqref{eq:linkprob} with $p_{ij}$ interpreted as deterministic values, not probabilities\footnote{Note that this may be referred to as a ``mean-field'' representation for the ensemble of networks for which $p_{ij}$ represents a probability.}.

\begin{figure}[t!]
  \centerline{
  \includegraphics[width=\columnwidth]{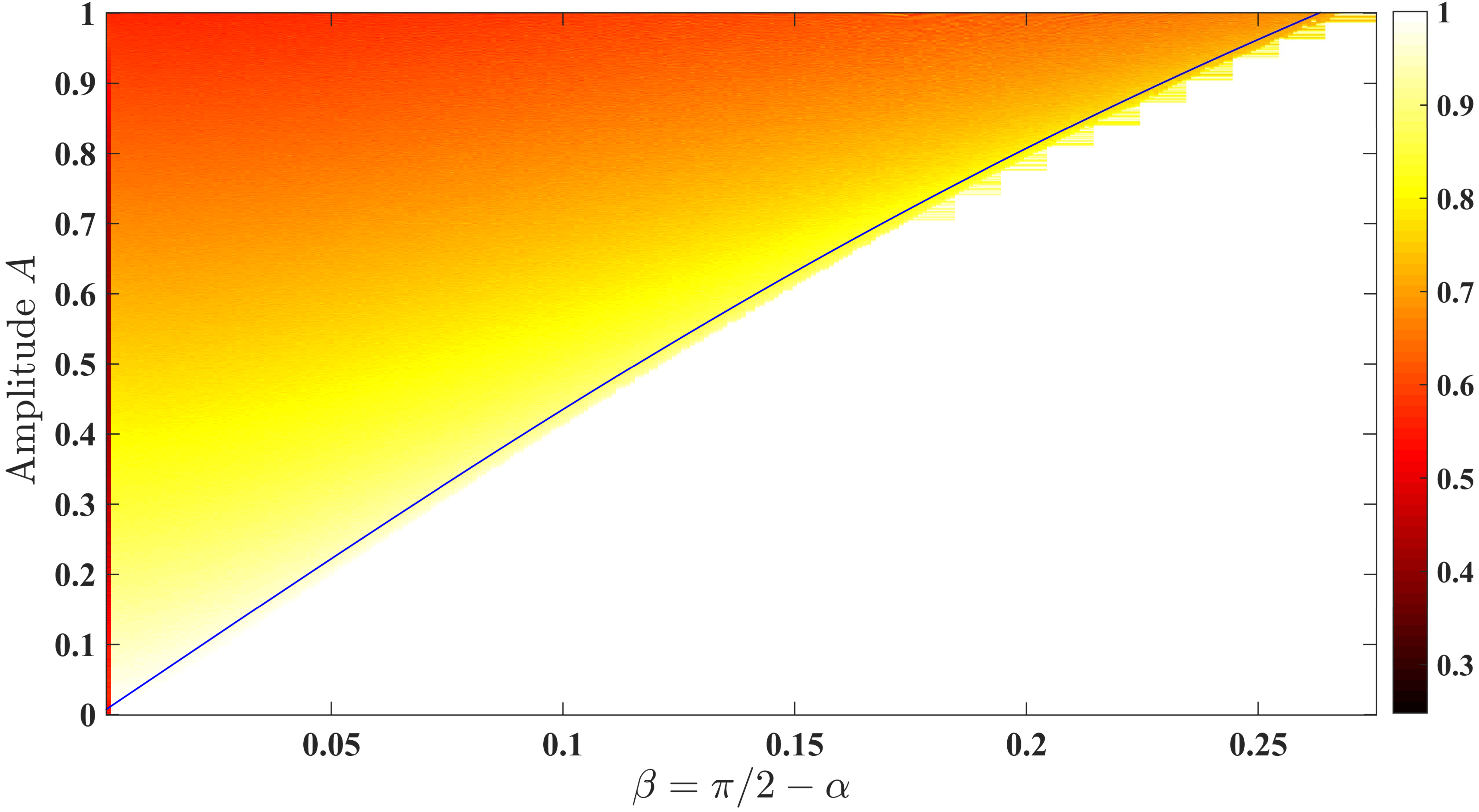}}
  \caption{(Color online) \textbf{Region of existence of chimera state.} Color indicates order parameter \eqref{eq:orderparam} in an all-to-all weighted network of coupled oscillators initialized in a chimera state.  Black line displays theoretical boundary where chimera state ceases to exist in continuum model.  Here $N=512$, $\lambda=\pi$. Numerical continuation was used and proceeded from $A=1$ down to $A=0$ for each fixed $\alpha$: see appendix for details on how initial conditions were generated for each $\alpha$.}
  \label{fig:vary_alpha_results}
\end{figure}

This experiment can be thought of as the closest numerical implementation of the continuum model when restricted to finite $N$. A detailed discussion of the connections between finite-size and continuum models is beyond the scope of this paper, but we refer the reader to the effective mean-field approach used in Ref.~\cite{Restrepo2005}. Unsurprisingly, our experiment resulted in a good match to theoretical predictions: see Fig.~\ref{fig:weighted_results} (data also displayed via the red dots in Fig.~\ref{fig:rewiring_results}) and Fig.~\ref{fig:vary_alpha_results}.

The experiments with all-to-all weighted networks served a similar function to those on binary irregular networks: they let us examine whether the idea of a partially synchronized chimera state extends to the case of finite $N$, though they did not allow us to explore the effects of irregularity and symmetry breaking.  Nevertheless, because of the greater ease with which all-to-all weighted networks can be simulated, we were able to do more extensive tests of the full parameter space of existence of chimera states, as shown in Fig.~\ref{fig:vary_alpha_results}.

\paragraph*{Discussion.}
In a sense, it is unsurprising that we found good agreement between continuum theory and experiment for both weighted and binary finite networks: in both cases, we have constructed discrete approximations of the continuum coupling kernel examined in \cite{Abrams2004}. Our experiments might even be thought of as examining a dynamical bifurcation onset brought about by slow variation of the parameter $A$, implemented via rewiring.

In another sense, however, it is quite surprising that these persistent partially synchronized states may exist on finite networks, where the symmetry of the state differs from the symmetry of the network (or more exactly, the symmetry of the expected value of the network structure).  Substantial recent research effort has gone into computing network symmetries and network motifs, with the (sometimes unstated) underlying assumption that these network properties will have implications for dynamics and equilibria on the network. It may be computationally impractical to find all possible symmetry groups for subsets of a large finite network, and our numerical experiments show that employing intuition drawn from the symmetry of the average connectivity will miss something.

Imagine our first numerical experiment being conducted in reverse, with structure gradually imposed on a random initial matrix.  At some point, though the spatially homogenous state remains stable, a new apparently stable heterogeneous state is born ``out of the clear blue sky,'' through what would be a saddle-node bifurcation of chimera states in the continuum theory.  It's unlikely that even the existence of this state would be noticed in simulations of such a system, where nothing in the symmetry suggests it.  If this occurred in an engineered system such as the US power grid, where global synchrony of generators is crucial, accidental entry into the partially-synchronized state could be catastrophic.

In our numerical simulations, we found that $N$ needed to be fairly large ($N \gtrsim 500$) to prevent stochastic alternation between the fully and partially synchronized states.  We also found that the network needed to be fairly dense ($\lambda \gtrsim 1$) for both states to coexist. This suggests that ``surprise'' coexistence---coexistence of symmetric and unintuitive symmetry-breaking persistent states---may only occur in larger and denser networks.

\paragraph*{Conclusions.}
A major open question in the study of complex networks is as follows: given a structured network, what stable dynamical states may exist?  In this paper we have shown that counter-intuitive states analogous to chimera states may coexist with intuitive spatially homogeneous states.  It is difficult to anticipate when to expect these, especially if node layout in physical space is not known or meaningful (e.g.~see lower panels of Fig.~\ref{fig:rewiring_matrix}).

It remains unknown if this chimera-state analogue is the only symmetry-breaking persistent state on the networks we analyze, or whether others might remain hidden in the extremely high dimensional phase space of a large network.

\appendix 
\section{Numerical notes} \label{app:num_notes}
We find initial conditions (ICs) for the chimera state in the following way.  Setting $\alpha=1.4$, we generate an IC by drawing uniform random phases $\phi_i \in (-\pi, \pi]$.  We evolve the dynamical system forward in time using an adaptive 4th/5th order Runge-Kutta method as implemented in the commercial Matlab software (version R2014a), and because the basin of attraction of the synchronous state is very small with this $\alpha$ value, the system typically spontaneously reaches the chimera state.

After equilibration in the chimera state, we use numerical continuation to gradually change $\alpha$ to the desired value (e.g., $1.34$ as in many of our figures) pseudostatically (i.e., we evolve the dynamical system to equilibrium after each small step in $\alpha$).  In the case of the binary network, the resulting distribution of $\phi$ values serves as the initial condition for each network realization in an ensemble.  In the case of the all-to-all weighted network, differing ICs are generated by integrating the dynamical for a large random time interval while holding $\alpha$ constant at the desired value.

\section{Rewiring as superposition} \label{app:rewiring}
The rewiring process discussed in the manuscript can be thought of as linear superposition of two objects: (1) the spatially-structured (initial) network described by an adjacency matrix with link probability according to Eq.~\eqref{eq:linkprob}, and (2) the Erd\"{o}s-R\'{e}nyi random (target) network where link probability is independent of spatial position.  In the continuum limit, the adjacency matrix for the initial network is replaced by the coupling kernel given in Eq.~\eqref{eq:chimerakernel}, and the adjacency matrix for the target network becomes independent of spatial position so that the effective coupling kernel is $G(x,x')=1/2 \pi$.


When the link rewiring experiment proceeds as described in the text, a new coupling matrix $\mathbf{C}$ is created at each step, combining some links from the initial network with others from the target network.  In an ensemble average sense, this is equivalent to a weighted sampling of links from the initial matrix $\mathbf{I}$ and target matrix $\mathbf{T}$.  In expected value $\mathbf{C}$ is just the weighted mean $\mathbf{C} = (1-w) \mathbf{I} + w\mathbf{T}$, where $w \in [0,1]$ is a weight equivalent to the relative amount of rewiring $M / \Mmax$.

The same idea applies rigorously in the continuum limit, where rewiring is equivalent to linear superposition of continuous coupling kernels.  Thus 
\begin{align*}
  G(x,x') &= (1-w) \left\{ \frac{1}{2\pi}[1+A \cos(|x-x'|)] \right\} 
  + w \left( \frac{1}{2 \pi} \right) \\
    &= \frac{1}{2\pi}[1+A' \cos(|x-x'|)],
\end{align*}  
where $A' = A(1-w)$ is the ``equivalent A'' resulting from the superposition.  This linear relationship between ``equivalent A'' and rewiring $w=M/\Mmax$ is the reason for the observed linearity in the right panel of Fig.~\ref{fig:equivalentA}.

Using different initial and target matrices (or, in the continuum limit, coupling kernels) would yield a different superposition that may or may not be as easily parameterized.

\section{Why a chimera state} \label{app:whatisit}

We refer to the partially synchronized state shown in the lower left panel of Fig.~\ref{fig:twostates}---the focus of this manuscript---as analogous to a ``chimera state.'' One might ask how it differs from the state shown in the lower right panel of Fig.~\ref{fig:twostates}, which also includes some asynchronous oscillators.  In particular, if relabeling of node spatial position is allowed, can the two states be viewed as interchangeable?

The answer is no: the two states are intrinsically different. The phenomenon that we find counter-intuitive is that identical oscillators on a network can enter a persistent partially-synchronized state which does not reflect the network symmetry (and as far as we are aware, the chimera state is the only persistent partially-synchronized state observed for networks of identically coupled identical phase oscillators in the $N \to \infty$ limit).

If the system is not in the ``chimera state'' analogue, simulation of Eq.~\eqref{eq:ode} with large $N$ should always produce something like the lower right panel of Fig.~\ref{fig:twostates}, where the phase and average frequency distributions do not agree with theory from \cite{Abrams2004}.  Additionally, the set of oscillators that are ``drifting'' should reflect particular symmetries of the coupling network, rather than resulting from the underlying symmetry-breaking of the chimera state.

\begin{acknowledgments}
The authors thank O.~E.~Omel'chenko and M.~Wolfrum for helpful discussions. This work is partially supported by NSFC grants No.~11201017 and 11290141.
\end{acknowledgments}

\bibliography{chimera_network_v3}

\end{document}